A Modifiable Architectural Design for Commercial Greenhouses Energy Economic Dispatch Testbed

Clausen, Christian Skafte Beck; Jørgensen, Bo Nørregaard; Ma, Zheng Grace

Download date: 08. Jan. 2024

# A Modifiable Architectural Design for Commercial Greenhouses Energy Economic Dispatch Testbed

Christian Skafte Beck Clausen [1[0000-0003-3118-7253]], Bo Nørregaard Jørgensen [1[0000-0001-5678-6602]] and Zheng Grace Ma [1[0000-0002-9134-1032]]

[1] SDU Center for Energy Informatics, Maersk Mc-Kinney Moeller Institute, The Faculty of Engineering, University of Southern Denmark, Odense, Denmark
`csbc@mmmi.sdu.dk`

**Abstract.** Facing economic challenges due to the diverse objectives of businesses, and consumers, commercial greenhouses strive to minimize energy costs while addressing $CO_2$ emissions. This scenario is intensified by rising energy costs and the global imperative to curtail $CO_2$ emissions. To address these dynamic economic challenges, this paper proposes an architectural design for an energy economic dispatch testbed for commercial greenhouses. Utilizing the Attribute-Driven Design method, core architectural components of a software-in-the-loop testbed are proposed which emphasizes modularity and careful consideration of the multi-objective optimization problem. This approach extends prior research by implementing a modular multi-objective optimization framework in Java. The results demonstrate the successful integration of the $CO_2$ reduction objective within the modular architecture with minimal effort. The multi-objective optimization output can also be employed to examine cost and $CO_2$ objectives, ultimately serving as a valuable decision-support tool. The novel testbed architecture and a modular approach can tackle the multi-objective optimization problem and enable commercial greenhouses to navigate the intricate landscape of energy cost and $CO_2$ emissions management.

## 1 Introduction

Commercial greenhouses confront economic challenges stemming from the dynamic objectives of businesses, consumers, and policymakers. From a business standpoint, greenhouses aim to minimize energy costs, while from a societal perspective, stakeholders such as consumers and policymakers strive to reduce $CO_2$ emissions. These challenges are intensified by escalating energy costs and the global urgency to curb $CO_2$ emissions. The landscape is further complicated by fluctuating energy prices and $CO_2$ emissions dependent on specific energy technologies. Commercial greenhouses operating in cold and low-light climates utilize heterogeneous energy systems to meet plant growth demands by providing supplemental heat and electricity. In the prevailing scenario, greenhouse operators determine daily optimal energy system operation based on fluctuating energy prices, climate data, and plant information. This economic dispatch procedure, typically executed in spreadsheets, is both costly and error-prone, involving models that are difficult to adapt to evolving business, consumer, and policy-maker objectives.

Therefore, to address these dynamic economic challenges, this paper aims to present an architectural design for a testbed focused on the modifiability of energy economic dispatch in commercial greenhouses, addressing the associated challenges.

The paper is organized as follows. Section 1.1 describes related research. Section 2 describes the methodology. Section 3 identifies stakeholders and high-level requirements. Section 4 describes the experimental setup. Section 5 presents the results. Section 6 discusses the results. Section 7 concludes the paper and lists future work.

### 1.1 Related Works

Economic dispatch is a regular method in energy systems that aims to dispatch instructions to generation units while minimizing costs given the constraints of load demands [1]. Costs include all relevant economic aspects in the domain. Economic dispatch has been applied in e.g., power systems [2-4], demand response [5], and greenhouses [6, 7]. In essence, economic dispatch is an optimization problem and can be implemented in numerous ways; merit order dispatch [8, 9], multi-objective optimization [2, 10], genetic algorithms [11-13], and particle swarm optimization [14] to mention a few. Each method has inherent advantages/disadvantages.

The economic dispatch algorithm must be verified before deploying it to operation. The verification methods apply the dispatch instructions on generation units in a simulated environment. The type of environment depends on the application's needs, but examples include discrete-event and real-time simulation environments. The purpose is to verify the performance of the dispatcher in open- or closed-loop control systems. The control loop requires models of the generation units under the system environment conditions. White-, grey-, and black-box methods are applied to deriving models that describe the reality of the systems. Recent approaches include digital twins that utilize the Internet of Things (IoT) to (i) mimic the behavior of the physical twins and (ii) operate the physical twin through its inherent cognitive abilities [15].

In-the-loop paradigms can utilize the virtual model or digital twin at various development stages. Software-in-the-loop (SIL) is applied at the early stage to assess the feasibility of the dispatch strategy [16, 17]. The benefit of SIL is that the control logic can be reused throughout the lifetime of the system. This enables rapid feedback cycles that lower costs and reduce risks before deploying the controller logic to production.

To manage and reduce the cost of software changes, flavors of this quality have been proposed: Modifiability, modularity, changeability, flexibility, adaptability, maintainability, and extensibility. Modifiability is a fundamental quality of software systems because changing requirements cause software evolution over the entire lifetime [18]. Therefore, software needs to be modifiable in places that are likely to change. The modifiability parameters involve modularization, increasing cohesion, and reducing coupling. These parameters can be realized to address changes at different points in time. For example, a change may be manifested during coding-, build-, deployment-,

initialization-, or runtime. The existing literature on economic dispatch focuses mostly on the optimization methodology rather than software modifiability and changing needs. The authors in [19], however, demonstrated a climate control application with the ability to add and remove independently developed objectives through a global lookup registry called a blackboard architecture. This paper contributes to the existing literature by suggesting an architectural testbed for economic dispatch for greenhouse energy systems. In contrast to the existing literature, this paper includes (i) an analysis of the architectural issues, (ii) integrates Functional Mock-up Units (FMU) to mimic individual energy production units, (iii) and demonstrates the consequence of various decision-making strategies among Pareto-optimal solutions.

This paper consequently uses the term modifiability to describe the degree to which the system can be changed (add/delete/modify) to meet new requirements [20]. This requires the identification of foreseeable changes through module decomposition which will reduce the change's cost impact.

## 2 Methodology

This paper applies software engineering principles to design the architecture for the energy economic dispatch in commercial greenhouses with a focus on modifiability. The approach follows Attribute-Driven Design (ADD) proposed by Bass et al. [21]. The goal of ADD is to propose an initial architecture that enables iterative refinement. This method is depicted in **Fig. 1.** The Attribute-Driven Design (ADD) methodology. The driver of ADD is Architectural Significant Requirements (ASRs) i.e., the important functional and non-functional requirements. The ASRs are derived from high-level stakeholder- and business requirements. These requirements are used to sketch the core architecture where solutions to each architectural element are designed in subsequent iterations. The design must be verified by using appropriate techniques. This paper verifies the design by using a generic energy system of a greenhouse. Experimentation with physical energy production systems exposes the business to significant risks. Therefore, this paper operates on virtual models of energy production systems.

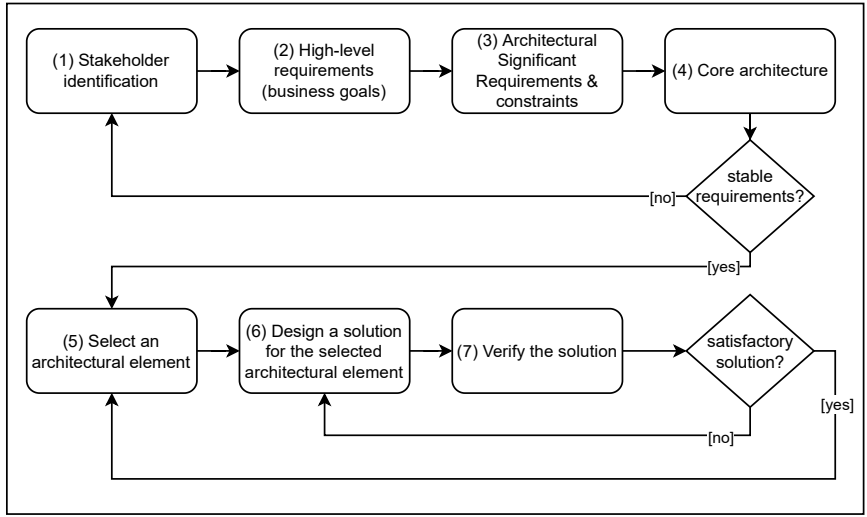

**Fig. 1.** The Attribute-Driven Design (ADD) methodology.

The paper extends previous research in [22, 23] and by applying a Java-based modular multi-objective optimization framework developed at the SDU Center for Energy Informatics [24]. This framework was developed with core design guidelines that focus on modularity. Other Multi-objective Evolutionary Algorithm (MOEA) frameworks are available e.g., jMetal, and MOEA Framework, but they do not satisfy the requirements of (i) providing decision making strategies to select ideal solutions from the Pareto-front and (ii) out-of-the-box integration with Functional Mock-up Interfaces (FMI) and Functional Mock-up Units (FMU). This includes the separation of algorithm configuration, decision variables, objectives, and decision-making strategies. Furthermore, it is important that the framework is interoperable with upstream and downstream systems to function with in-the-loop control approaches.

The chosen framework provides a multi-objective genetic algorithm (MOGA) The MOGA is an evolutionary heuristic search algorithm that progresses in iterations. In the first iteration, a population of candidate solutions is randomly generated. Each solution is evaluated against the set of objectives and the evaluation yields metrics to compare the optimality of a solution against other solutions. Since objectives may conflict, the currently best solutions found by the MOGA are stored in a Pareto-optimal set. In the subsequent iterations, the MOGA progresses by selecting parent solutions from the Pareto-set. Based on the parents, new candidates are created by applying the MOGA operators of crossover and mutation. Crossover selects genes from each parent, and mutation alters the value of a gene to a random value to diversify the population. Crossover also enables the MOGA to escape local minima. The MOGA terminates upon reaching a termination criterion which in this case is temporally determined. MOGA provides the advantages of being intuitive to understand, fast to implement, and enabling separate modification of objectives during compilation- and runtime. The modification of multiple objectives is possible because the objective does not interfere.

55

This property enables an architecture where objectives can be plugged in/out. The significant disadvantage of MOGA is the high computational cost. This is an important factor for real-time applications, and the application designer must decide whether the computational cost can be afforded. For this paper, the computational window for decision-making is at least 10 minutes. This window is dictated by the control of the physical equipment. Furthermore, greenhouse climate operators prefer slow ramp-ups/downs of their equipment because it is empirically cost-efficient. This gives time for the MOGA to converge towards optimality.

The paper applies Functional Mock-up Units (FMU) for modeling the energy system dynamics. An FMU is an executable unit that encapsulates the design and behavior of a system as a black box simulation. FMUs provide the necessary interfaces and code to exchange dynamic simulation models of the greenhouse's energy systems. The FMUs in this paper were developed in Dymola/Modelica.

## 3     High-level Stakeholders' Requirements Identification

The stakeholders and their interests were identified through coordinated meetings and seminars. The result is shown in **Table 1**. The stakeholders and interests are important factors in scoping the architecture through ASRs.

**Table 1.** Stakeholders, descriptions, and their interests.

| Stakeholder | Description | Concerns |
| --- | --- | --- |
| Greenhouse business owner | Owns the greenhouse and is therefore responsible for achieving its overall goal. | To meet the business goals and the economic aspects of: <br>• Maximizing revenue. <br>• Minimizing expenses. <br>• Complying with regulatory requirements. |
| Greenhouse grower | The grower is responsible for nurturing the plants including (1) watering, (2) applying fertilizer and retardation, (3) moving plants between conveyors, and (4) logging events for regulatory authorities. | • To comply with the expected plant quality (size, number of leaves, etc.). |
| Greenhouse climate operator | The greenhouse climate operator is responsible for creating ideal growing conditions by monitoring and controlling the indoor environment. For example, artificial lights, curtains, temperature, and $CO_2$. | • To operate the indoor climate while complying with the growing conditions of the plants. |
| Customer | The customer buys the plants. | • To buy plants of the expected quality and price. |
| Policy maker | Makes international, national, or local policies that are motivated by social and political factors. | • To make and enforce regulations. |
| External data provider | The external data provider is responsible for providing services or data that support the greenhouse business operation. | • To provide relevant data of a certain quality (including resolution) |
| | | that can help the growers in the decision-making process. |
| Energy Systems Engineer | The energy systems engineer provides the virtual models to simulate the dynamics of the greenhouse energy systems. | • To provide virtual models that mimic the real-world properties of the greenhouse energy systems. |
| Systems Engineer | The systems engineer manages and digitalizes the greenhouse business processes to help the business owner achieve their goals. | To develop a system that: <br>• Meets the expectations of all stakeholders. <br>• To minimize the development effort by managing complexity. <br>• To enable rapid development cycles to meet the need for frequent changes in the greenhouse operation. |
| Researcher | The role of the researcher is to advance research within the optimization of energy systems. | To have an architectural testbed that enables experimentation through: <br>• Crisp interfaces and abstractions. <br>• Analysis of algorithmic output. <br>• Translation of business goals into algorithmic objectives. |

The use of the system is two-folded since the system aims to address research-related- and business-related challenges. Ideally, the energy dispatch algorithms developed on the architectural testbed can be effortlessly deployed on the real system. **Fig. 2** shows a high-level development process of this ideal. The architecture must support the software-in-the-loop paradigm by operating on virtual models that comply with the interfaces and behavior of the real system. To realize this, the researcher, systems engineer, and energy systems engineer collaborate to realize the control software during early development in stage 1. The control software operates in a closed-loop mode to verify that the controller operates as expected. In stage 2, the virtual models are substituted with a digital twin of the energy systems. This enables verification of the real-time properties of the closed-loop controller. Furthermore, it enables the climate operator to test the decision support tools using the digital twin. Due to high investment costs and physical space requirements, it is infeasible to test the control software on actual hardware using hardware-in-the-loop in a lab. Testing on a digital twin lowers investment costs and enables rapid development cycles. When the test in Stage 2 is satisfactory, the controller software is configured to operate on the real hardware in Stage 3.

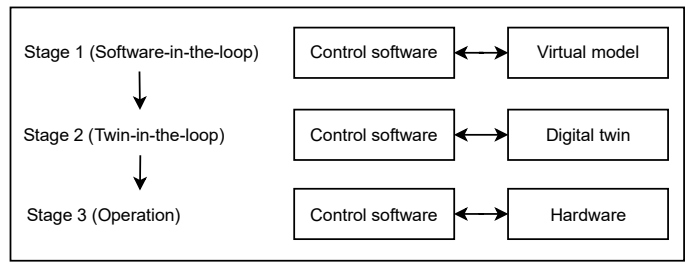

**Fig. 2.** Development stages using software-in-the-loop and digital twins in process control.

The users of the architectural testbed are researchers and climate operators. The researcher's use case is to utilize the architectural testbed to provide a feasible method of economic dispatch. The climate operator use case is decision-making support by using the output of the economic dispatch algorithm to operate the energy systems. Based on the stakeholder analysis and verification requirements, a set of high-level requirements was formulated as presented in **Table 2**.

**Table 2.** High-level system requirements.

| ID | Description | Requirement type |
|---|---|---|
| Requirement 1 | The system must address the economic dispatch problem by providing tradeoffs between conflicting business objectives. | Functional |
| Requirement 2 | The system must provide monitoring, scheduling, and decision-making features to assist the climate operator in decision-making. | Functional |
| Requirement 3 | The system must provide a testbed for developing economic dispatch control strategies. This includes the possibility of algorithmic performance evaluation. | Functional |
| Requirement 4 | The system must be modifiable to address the challenge of research-based activities and continuous change throughout the greenhouse operation's lifetime. | Non-functional |
| Requirement 5 | The system must support interoperability with test facilities and existing infrastructure e.g., forecasts services, virtual models, digital twins, and SCADA systems. | Non-functional |

### 3.1 Architectural Design

Following the ADD method, this section presents a high-level view of the core system architecture (**Fig. 3**) and subsequently narrows the focus to the economic dispatch optimizer. The design intends to satisfy the requirements by applying the process-control [25, 26] and in-the-loop paradigms. Therefore, the core design must hold throughout the three development stages.

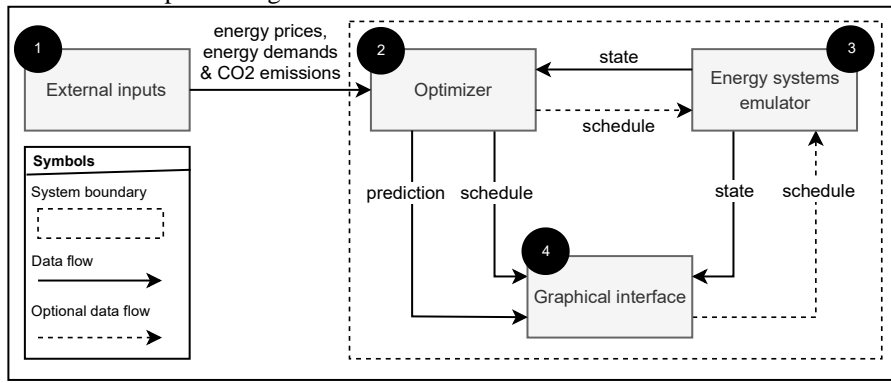

**Fig. 3.** Architectural design of the economic dispatch testbed.

The architecture is decomposed into four subsystems: 1) External inputs, 2) an optimizer, 3) an emulator, and 4) a graphical interface. This decomposition draws a natural boundary between the external data providers and the internal system. Furthermore, the architecture enables flexibility, and design choices can be delayed until the refinement of each subsystem. From a functional perspective, it gives the following properties:

- The optimizer isolates the economic dispatch optimization problem to accommodate experimentation with control strategies and performance evaluation. Furthermore, the optimizer can be executed on dedicated computational resources to improve performance using horizontal or vertical scaling.
- The emulator exposes a black box process with controls and the state of the energy system. The emulator can be developed independently by an energy systems engineer in any environment. This assumes that the optimizer and emulator models are interoperable. Furthermore, the emulator can be substituted by the real cyber-physical system.
- The graphical interface can be used by the climate operator for monitoring and decision support purposes. The exact usability requirements and technicalities can be delayed. But for example, it enables isolated development, and optional programming environment or thin clients for mobile units.

The execution of the economic dispatch algorithm is described on a high level as follows:
1. The optimizer is instantiated with a configuration of the multi-objective optimization problem. This includes forecast inputs and the state of the energy system.
2. The optimizer executes until a termination criterion is met. The output is a vector that consists of setpoints for each energy system in a certain resolution (hourly, minutely, …). Essentially, this is a prescription intended to control the energy systems and to inform the climate operator.
3. The optimizer sends the schedule (the current set of setpoints) and the prediction (the predicted result of actuating the setpoints) to the graphical interface. Optionally, the optimizer can send the schedule to the emulator when operating in closed-loop mode.
4. The climate operator uses the graphical interface to monitor the state, schedule, and predicted behavior of the energy systems. Furthermore, the climate operator can choose to manually actuate the schedule if running in human-in-the-loop mode.

From a non-functional perspective, the design aims to satisfy modifiability and interoperability at appropriate system levels (Requirements 4 and 5). Modifiability involves the identification and management of foreseeable changes. Interoperability involves information exchange between systems. On a system level, changes are localized to their respective subsystem assuming stable subsystem interfaces. Changes may cascade if the interoperability assumptions are wrong, i.e., data, sequencing, timing, endpoints, and pre-and post-conditions. Changes regarding interoperability within the boundary are manageable, but unforeseeable changes to the external inputs expose the architecture to wrong assumptions. This risk can be managed by accommodating interoperability through a modifiable architecture.

In the next step of the ADD, the internal structure of the optimizer (**Fig. 4**) is analyzed since this is the central subsystem of the architecture. The structure reflects the abstractions that need to be modifiable. Most abstractions are provided by the chosen

MOEA framework [24] and need only to be specialized in subclasses. The 'External Inputs' module is responsible for upstream interoperability by fetching forecasts and states and making that data compatible with the framework. The 'Decision maker' module provides a configuration of how the 'best' decision from the Pareto-frontier must be selected. This can be used to select the setpoints in the closed-loop scenario with a Digital twin. The 'Output' module is responsible for downstream interoperability with the digital twin and the graphical interface.

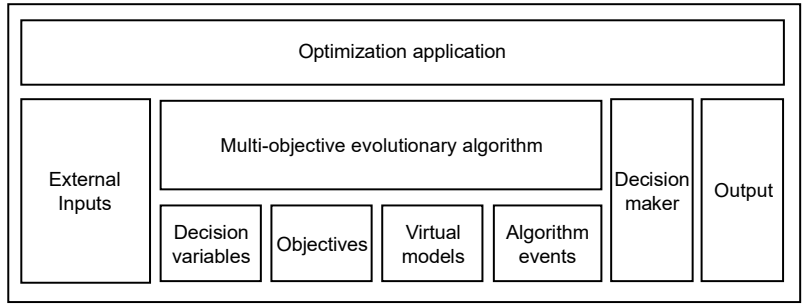

**Fig. 4.** Internal structure of the optimization application

The 'Multi-objective evolutionary algorithm' module consists of specialized decision variables, objectives, and the algorithm provided by the framework. Decision variables involve the encoding/decoding of the problem. The decision variables e.g., resolution, length, etc. are localized here. The objectives are responsible for evaluating the decision variables into scalar values using external input and the estimated response on the integrated virtual models. The virtual models are integrated into the optimizer as individual components (i.e., heat pumps, combined heat and power plant, gas boiler, district heating, and thermal energy storage). This design accommodates changes in the greenhouse energy system because the component composition can be reflected in code. For example, it can be used to experiment with up- or down-scaling or with heterogeneous energy production technologies. The framework emits algorithmic events that can be used to collect data for performance evaluation e.g., on each generation or upon termination.

## 4 Experimental Setup

The purpose of the experiment is to demonstrate the economic dispatch in a hypothetical greenhouse application built upon the modular architecture. The experiment includes the following setup:

### 4.1 Greenhouse Energy System Configuration

The case is based on a hypothetical greenhouse compartment with heterogeneous energy systems. The compartment's energy system was composed of virtual models and a block diagram of their relations is depicted in **Fig. 5** where the corresponding symbols are defined in **Table 3**. The models were implemented in Dymola/Modelica and the models were exported to FMUs for integration with the optimizer's objective functions. The virtual models in this paper were derived with grey-box modeling techniques using on-site measurements from a commercial greenhouse in Denmark. White-box modeling is computationally expensive, and black-box modeling does not capture the physical properties of the system. Furthermore, black-box modeling relies on high-quality training data which were unavailable. The grey-box models approximate the system response using the collected measurements to tune the parameters of theoretical models. This approach is suitable for this paper where the focus is the architectural design and because the model precision can be tuned later. The system dynamics were simplified to linear equations and the thermal energy storage does not account for temperature gradients or heat loss.

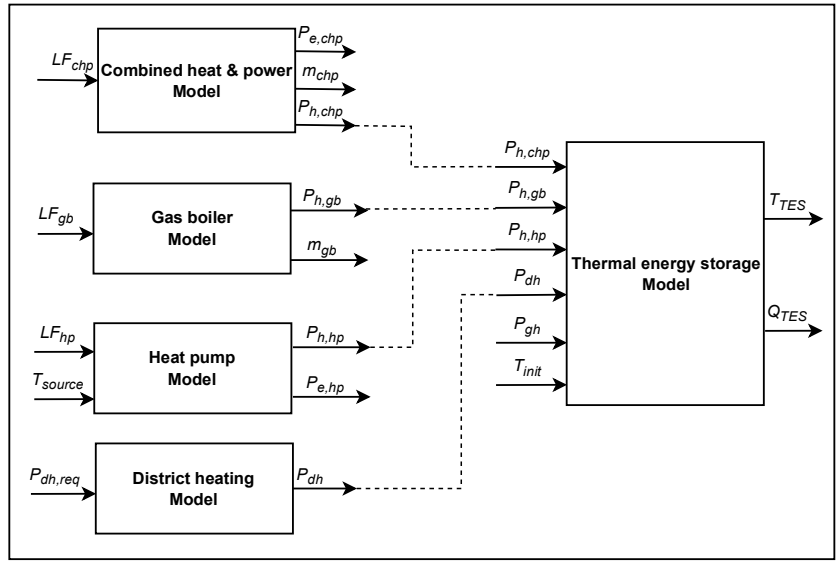

**Fig. 5.** Block diagram of the experimental greenhouse compartment's energy system.

**Table 3.** Block diagram symbols and descriptions.

| Symbol | Unit | Interval | Description |
|---|---|---|---|
| $LF_{chp}$ | [-] | [0, 1] | Load factor of the combined heat & power (CHP) model. |
| $LF_{gb}$ | [-] | [0, 1] | Load factor of the gas boiler (GB) model. |
| $LF_{hp}$ | [-] | {0, 1} (binary) | Load factor of the heat pump (HP) model. Either on or off. |
| $T_{source}$ | [°C] | [20, 50] | Water temperature at the inlet of the evaporator of the HP. |
| $P_{h,chp}$ | [MW] | [0, 2.8] | Heat power produced by the CHP. |
| $P_{h,hp}$ | [kW] | [0, 500] | Heat power produced by the HP. |
| $P_{h,gb}$ | [MW] | [0, 7] | Heat power produced by the GB. |
| $P_{dh,req}$ | [MW] | [0, 6] | The district heating (DH) power request from the DH network. |
| $P_{dh}$ | [MW] | [0, 6] | Obtained district heating power from the DH network. |
| $P_{gh}$ | [MW] | [0, 10] | Greenhouse heat demand. |
| $T_{init}$ | [°C] | [0, 90] | The initial temperature of the thermal energy storage (TES). |
| $P_{e,chp}$ | [MW] | [0, 1.2] | Electric power produced by the CHP. |
| $m_{chp}$ | [kg/s] | [0, 12] | Gas mass flow rate consumed by the CHP. |
| mgb | [kg/s] | [0, 194] | Gas flow rate consumed by the GB. |
| Pe,hp | [kW] | [0, 125] | Electric power consumed by the HP. |
| $T_{TES}$ | [°C] | [0, 90] | Water temperature of the TES. |
| $Q_{TES}$ | [MWh] | [0, 65] | Stored energy in the TES. |

The input vector used to instantiate the energy system is defined by

$$S_{init} = [LF_{chp}, LF_{gb}, LF_{hp}, T_{source}, T_{init}, P_{dh,req}, P_{gh}] \tag{1}$$

where the output of each model is calculated. The heat output vector $P_{out} = [P_{h,chp}, P_{h,gb}, P_{h,hp}, P_{dh}]$ is used as input parameters of the TES. The TES operates analogous to a battery that can be charged and discharged but with heat instead of electricity. The TES enables flexible operation because heat can be stored when prices are low and drained when prices are high. The outputs $E_{out} = [P_{e,chp}, m_{chp}, m_{gb}, P_{e,hp}, P_{dh}]$ are used to calculate the economics of the energy system in terms of costs and $CO_2$ emissions.

### 4.2  Multi-objective Problem Definition

The energy dispatch schedule is represented as a decision vector that contains the energy systems load factors for each instant in the schedule:

$$S = [C_i, C_{i+1}, \ldots, C_{i+h}, G_i, G_{i+1}, \ldots, G_{i+h}, H_i, H_{i+1}, \ldots, H_{i+h}, D_i, D_{i+1}, \ldots, D_{i+h}]^T \tag{2}$$

where $S$ is the schedule, $i$ is an instant, $C$ is the CHP load factor, $G$ is the GB load factor, $H$ is the HP load factor and $D$ is the heat request to the DH provider. The number of instants $i$ are configured depending on the application requirements.

The optimization problem is subject to minimizing the costs and $CO_2$ emissions of the schedule:

$$\min O_{cost}(S) = \sum_{i=1}^{h} cost_{gas,i} + cost_{el,i} + cost_{dh,i} - income_{el,i} \tag{3}$$

$$\min O_{CO2}(S) = \sum_{i=1}^{h} CO2_{gas,i} + CO2_{el,i} + CO2_{dh,i} \tag{4}$$

where $h$ is the number of instants within the schedule. $cost_{gas,i}$ is the price of gas consumed in the instant, $cost_{el,i}$ is the price of electricity consumed in the instant, and $cost_{dh,i}$ is the price of district heating consumed in the instant. The CHP produces electricity which needs to be deducted from $income_{el,i}$ in the instant. $CO2_{gas,i}$ is the emissions of gas consumed by the CHP and GB. $CO2_{el,i}$ is the emissions of electricity consumed by the HP. $CO2_{dh,i}$ is the emissions caused by district heating.

The optimization is subject to the following objective space constraints:

$$C_{electricity}(S) = \forall i \in S : P_{gh,i} + P_{e,hp,i} \leq P_{grid\_capacity} \tag{5}$$

$$C_{TES}(S) = \forall i \in S : T_{TES,min} \leq T_{TES,i} \leq T_{TES,max} \tag{6}$$

$C_{electricity}(S)$ states that the greenhouse power demand and the HP power consumption must not exceed the grid capacity in any instant. $C_{TES}(S)$ states that the temperature of the TES must be within the minimum and maximum values in any instant.

The optimization is subject to the following decision space constraints to reduce the size of the decision space.

$$C_{CHP}(S) = \forall C_i \in S : (0 \leq C_i \leq 1) \wedge (C_i \bmod C_r = 0) \tag{7}$$

$$C_{GB}(S) = \forall G_i \in S : (0 \leq G_i \leq 1) \wedge (G_i \bmod G_r = 0) \tag{8}$$

$$C_{DH}(S) = \forall D_i \in S : (0 \leq D_i \leq D_{max}) \wedge (D_i \bmod D_r = 0) \tag{9}$$

$$C_{HP}(S) = \forall H_i \in S : H_i = 0 \vee H_i = 1 \tag{10}$$

$C_{CHP}(S)$ states that the load factor $C_i$ must be between 0 and 1 while being divisible by the resolution $C_r$. $C_{GB}(S)$ and $C_{DH}(S)$ have similar properties but with separate resolutions $G_r$ and $D_r$. $C_{HP}$ state that the load factor $H_i$ must be exactly 0 (turned off) or 1 (turned on).

### 4.3 Optimization Configuration

**Table 4** shows the parameters of the optimizer. The TES was configured with an initial temperature $T_{init}$ to enable immediate flexible use. For the experiment, it must not be economically feasible to drain the TES completely in the last instants of the schedule as the optimizer would regard the initial TES contents as free energy. Therefore, an additional objective constraint was created to avoid this case:

$$C_{TES\_end\_temperature}(S) = \forall i \in S : T_{init} \leq T_{TES,h} \tag{11}$$

where $T_{init}$ is the initial temperature of the TES and $T_{TES,h}$ is the temperature in the last instant.

**Table 4.** Optimization parameters.

| Scope | Parameter | Value | Unit |
|---|---|---|---|
| MOGA | Schedule length | 168 | [h] |
|  | Termination criteria | 1 | [h] |
|  | Crossover rate | 50 | [%] |
|  | Mutation operator | Random resetting | [-] |
|  | Crossover operator | Single-point crossover | [-] |
|  | Mutation rate | 5 | [%] |
| $C_{electricity}(S)$ | $P_{grid\_capacity}$ | $12 \cdot 10^6$ | [W] |
|  | Electricity Tariff and Fee | 0.185 | [EUR/kWh] |
| $C_{TES}(S)$ | $T_{TES,min}$ | 43.96 | [°C] |
|  | $T_{TES,max}$ | 79.84 | [°C] |
| $C_{CHP}(S)$ | $C_r$ | $\frac{1}{20}$ | [-] |
| $C_{GB}(S)$ | $G_r$ | $\frac{1}{20}$ | [-] |
| $C_{DH}(S)$ | $D_r$ | $10^4$ | [-] |
|  | $D_{max}$ | $6 \cdot 10^6$ | [W] |
| TES Model | $T_{init}$ | 50 | [°C] |
| HP Model | $T_{source}$ | 20 | [°C] |

### 4.4 External Inputs

Given the architectural boundary, it is assumed that the external inputs are available as forecasts. The forecasts consist of the greenhouse energy demands (electricity and heat), electricity-, gas-, and district heating prices, and $CO_2$ emissions for each technology. The forecasted greenhouse demands are depicted in **Fig. 6**. The forecasted energy demands are based on optimal climate and historical energy consumption records [23] The electricity wholesale price signal was retrieved from Nordpool in the Danish area DK1 and tariffs were retrieved from Energinet. The gas price signal was retrieved from the next day's index of the Exchange Transfer Facility (ETF) at the European Energy Exchange (EEX). The energy prices are depicted in **Fig. 7**. The $CO_2$ emissions for electricity are available in a 5-minute resolution at Energinet. The 5-minute resolution was aggregated into an hourly resolution to satisfy the optimization constraints. $CO_2$ emissions for district heating are declared as a yearly average and were retrieved from the Danish provider Fjernvarme Fyn. The $CO_2$ emissions for gas are estimated to be 204 kg $CO_2$ / MWh by the Danish Energy Agency. The estimated $CO_2$ emissions are depicted in **Fig. 8**.

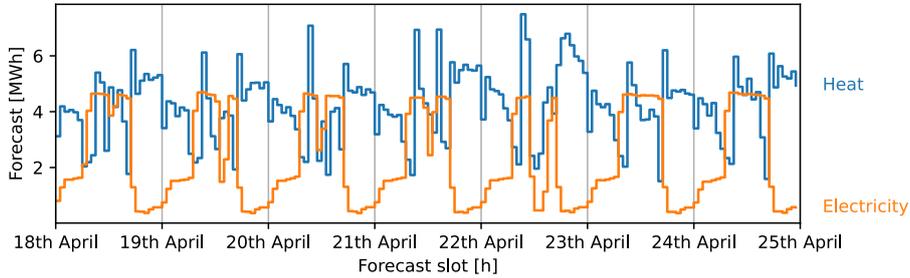

**Fig. 6.** Greenhouse heat and electricity forecast.

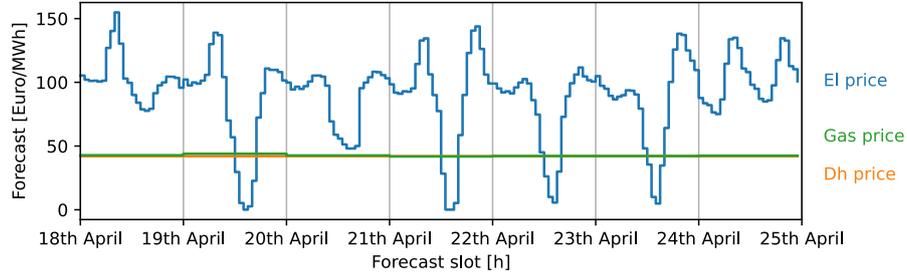

**Fig. 7.** Energy price forecast.

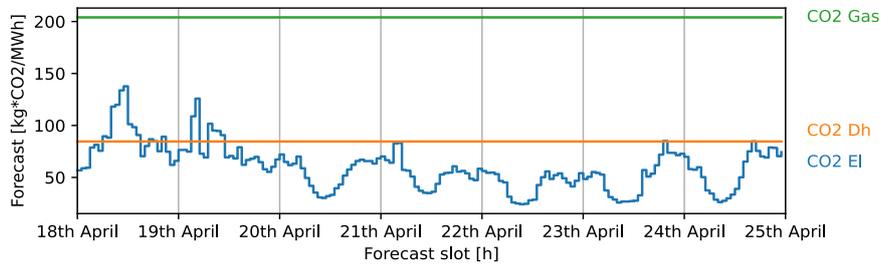

**Fig. 8.** Estimated $CO_2$ emissions forecast.

## 5 Results

The results were conducted on a workstation with an AMD Ryzen 9 5900X 12-core (24 threads) processor, 64 GB DDR4-3200 RAM, NVIDIA GeForce RTX 3060, and a Samsung 970 EVO NVMe drive. The termination criterion was set to 1 hour to match the lowest sample resolution from the external inputs. The data was exported in CSV format by using the built-in framework interfaces `GenerationFinishedListener` and `TerminationListener`.

**Fig. 9** depicts the resulting Pareto-front as a scatter plot of the objective space with $CO_2$ emissions [kg*$CO_2$] on the x-axis and costs [EUR] on the y-axis. The gray dots are invalid solutions while the red dots are valid solutions.

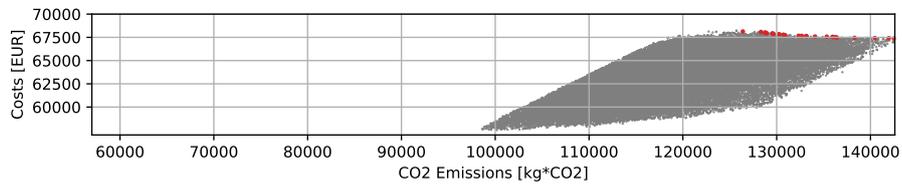

**Fig. 9.** Pareto-front in objective space; $CO_2$ emissions [kg*$CO_2$] vs. costs [EUR].

A solution is valid if the constraints of $C_{electricity}(S)$, $C_{TES}(S)$, and $C_{TES\_end\_temperature}(S)$ are satisfied. The number of invalid solutions is caused by the implementations of the constraints. To guide the MOGA in the objective space, the constraints were implemented as objectives that return a scalar value of the distance to each constraint. The distance is the sum of hours of unsatisfied electricity and heat, and the Euclidean distance to the target TES temperature.

The plot indicates an almost linear relationship between kg*$CO_2$ and costs indicating $CO_2$ emissions can be reduced without significantly increasing the costs. This can be explained by the constant $CO_2$ emission factors [kg*$CO_2$/MWh] of gas and district heating. Naturally, an increased energy consumption leads to increased $CO_2$ emissions. This explanation holds true for the gas and district heating $CO_2$ emissions signals because they are constant. However, $CO_2$ emissions for electricity consumption does not have this linear relationship, and therefore, the total emission depends on the fraction of energy produced by renewable energy sources with zero $CO_2$ emissions.

The resulting Pareto-front can be used for automated/manual decision-making. For the automated approach, the optimizer selects one solution by prioritizing or balancing costs or $CO_2$ using the built-in `DecisionMaker` interface. The result can also be sent to the climate operator for human-in-the-loop decision-making.

Suppose that the operator needs to choose an appropriate solution. **Fig. 10** shows three possible decision strategies: Minimize $CO_2$, minimize costs or make a compromise. This can be achieved by configuring the decision-making strategy by first removing the invalid solutions. Then, following the approach suggested in [27], all objective scores are normalized for comparison and have a social welfare metric applied. The minimum $CO_2$ and cost objectives are computed using an elitist metric, while the compromise is computed using the utilitarian metric. Simply put, the elitist metric selects the lowest value in either dimension. The utilitarian metric sums the normalized objectives scores and selects the solution with the overall lowest value.

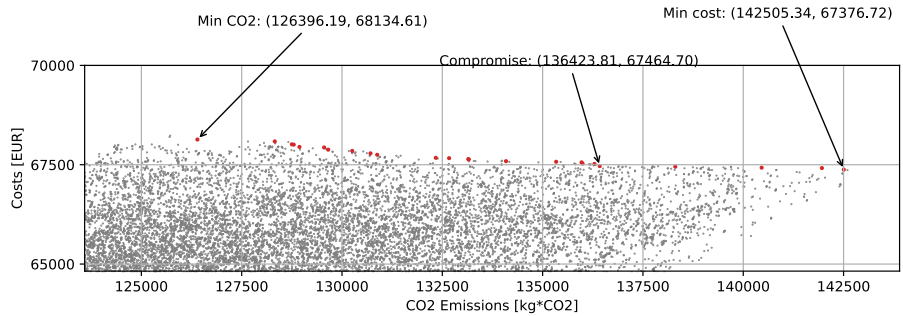

**Fig. 10.** Decision-making strategies within the Pareto-front.

**Table 5** shows a comparison of the three strategies. In the given experiment it is possible to reduce the costs by 757.89 [EUR] (1.11%) but this increases $CO_2$ emissions by 16109,15 [kg*$CO_2$] (12.74%) over a 168-hour period. 16 tons $CO_2$ corresponds approximately to the average $CO_2$ emissions of one US citizen in a year.

Table 5. Decision-making strategy comparison.

| Strategy | $CO_2$ [kg*$CO_2$] | Costs [EUR] | $CO_2$ index | Costs index |
|---|---|---|---|---|
| 1 | 126,396.19 | 681,34.61 | 100 | 100 |
| 2 | 136,423.81 | 674,64.70 | 107.93 | 99.02 |
| 3 | 142,505.34 | 673,76.72 | 112.74 | 98.89 |

## 6  Discussion

The paper extended the research in [24] by utilizing the modifiable architecture. The $CO_2$ objective was integrated by adding a new objective that operated on the original set of decision variables in the greenhouse domain. Changes to the $CO_2$ objective (add/remove/modify) are localized and do not interfere with the cost objective. The external inputs for $CO_2$ emissions were similarly added to the modular architecture by adhering to the hourly resolution in the domain. The $CO_2$ objective is limited in a few ways. Firstly, the district heating estimated $CO_2$ emission is a yearly constant. The objective space would change significantly if the estimates were more precise. Furthermore, there is a risk that averaging the 5-minute $CO_2$ input to an hourly rate skews the optimization results. For example, the $CO_2$ emissions might increase tremendously in a 5-minute period but decrease for the remaining hour. This information is lost when averaging. Secondly, the computational performance can suffer if the resolution is increased from hourly to 5-minute intervals, because the solution space increases drastically. Therefore, it is important to adjust the optimization resolution for real-world application needs. Thirdly, the FMUs could have a localized model that reflected $CO_2$ emissions more precisely depending on the efficiency of the production unit. The computational performance of the algorithm is limited by the FMU integration. FMUs consist of native C-code that must be called from Java when a decision vector is evaluated in the objective. To speed up the performance, a local cache was implemented. The cache stores the computed FMU output on a given decision vector. There is a high probability that equivalent decision vectors are generated due to the stochastic nature of MOEA. Tests showed that implementing the FMUs as linear functions could speed up the objective evaluation by a factor of 50. This computational cost associated with FMU integration is a serious trade-off consideration. A few limitations within the experiment are addressed next. There is a lack of demonstrating modifiability in the decision vector. Changes to the configuration of energy production units require the decision vectors to reflect these changes. There is a tight coupling between the decision vectors, objectives, and FMUs that must be addressed to lower modification costs further. Ideally, these modifiability changes could be made by the climate operator in a domain-specific language or user interface that focuses on usability. The results in hourly resolution should also be compared with decision-making frequencies that reflect the greenhouse domain, for example, to schedule the operation daily or weekly.

## 7 Conclusion and Future Work

This paper proposes an architectural design for a testbed for energy economic dispatch in commercial greenhouses. This includes the identification of stakeholders and their requirements along with an architectural design. The architecture demonstrated modifiability to the extent that the incoming change request for including $CO_2$ emissions in the decision-making process can be supported. Furthermore, the architecture separates the core modules in the optimizer to enable economic dispatch experimentation. For example, to experiment with other decision-making strategies, tuning, etc. The architecture aims to accommodate the foreseeable requirements of in-the-loop testing by applying the process-control paradigm. The paradigm separates the decision variables and optimization from the controlled process. This separation promotes modifiability and reusability of the controller in later in-the-loop stages. The proposed novel testbed architecture and a modular approach can tackle the multi-objective optimization problem and enable commercial greenhouses to navigate the intricate landscape of energy cost and $CO_2$ emissions management. Future work might consider (i) including a demonstration of modifiability in the decision vector and lowering the inherent coupling to the objectives. (ii) Assessing the costs of reconfiguring the energy system by adding new production units or removing obsolete units. (iii) Identifying more operation objectives such as favoring certain production units for impairment or maintenance reasons. (iv) The real-time capability of the core architecture should be tested at the subsequent development stages. Furthermore, (v) the demonstration must involve the graphical decision-making support tool and the digital twin of the greenhouse energy system.

## Acknowledgements

This paper is part of the IEA-IETS Annex XVIII: Digitization, artificial intelligence and related technologies for energy efficiency and reduction of greenhouse gas emissions in industry project, funded by EUDP (project number: 134-21010).